\documentstyle[12pt]{article}
\begin{document}
%%%%%%%%%%%%%%%%%%%%%%%%%%%%%%%%%%%%%%%%%%%%%%%%%%%%%%%%%%%%%%%%%%%%%%%%%%%%%%
\centerline{\bf THE NUMBER OF TENSOR DIFFERENTIAL INVARIANTS}

\centerline{\bf OF A RIEMANNIAN METRIC}

\bigskip\bigskip\bigskip

\centerline{{\bf Victor Tapia}\footnote{\tt TAPIENS@CIENCIAS.UNAL.EDU.CO}}

\bigskip

\centerline{\it Departamento de F{\'\i}sica}

\centerline{\it Universidad Nacional de Colombia}

\centerline{\it Bogot\'a, Colombia}

\bigskip\bigskip

\centerline{\bf Abstract}

\bigskip

We determine the number of functionally independent components of tensors involving higher--order derivatives of a Riemannian metric.

%%%%%%%%%%%%%%%%%%%%%%%%%%%%%%%%%%%%%%%%%%%%%%%%%%%%%%%%%%%%%%%%%%%%%%%%%%%%%%
\bigskip\bigskip

A central problem in Riemannian geometry is to determine the number of functionally independent invariants which can be constructed out from a Riemannian metric and its derivatives up to a given order. Due to the inmediate applications of these invariants as Lagrangians in the formulation of alternative gravitational theories, the work have been mainly oriented to scalar invariants \cite{1}--\cite{5}.

In this work we determine the number of functionally independent components of tensors involving higher--order derivatives of a Riemannian metric. As a starting point we adopt the taxonomic \cite{6} definition of a tensor based on the transformation rule of its components: {\it a tensor is something which transforms like a tensor}. This means that in the corresponding transformation rule only the transformation matrix, ${X^\mu}_\alpha=\partial x^\mu/\partial y^\alpha$ appears. When considering derivatives of the metric, up to a certain order $d$, derivatives of the transformation matrix will appear. By a simple counting it is possible to determine the number of relations not involving derivatives of the transformation matrix. This number turns to be the number of components of a tensor, of rank $(d+2)$, involving derivatives of the metric.

Let us now consider in detail the determination of the number of independent components of tensors involving derivatives of the metric. As announced, we will consider the transformation rule of the different objects. An $r$--rank covariant tensor $T_{\mu_1\cdots\mu_r}$ is an object which transforms like

\begin{equation}
T_{\alpha_1\cdots\alpha_r}({\bf y})={X^{\mu_1}}_{\alpha_1}\,\cdots\,{X^{\mu_r}}_{\alpha_r}\,
T_{\mu_1\cdots\mu_r}({\bf x})\,,\label{1}
\end{equation}

\noindent where

\begin{equation}
{X^\mu}_\alpha={{\partial x^\mu}\over{\partial y^\alpha}}\,.\label{2}
\end{equation}

\noindent For later convenience let us also introduce 

\begin{equation}
{X^\mu}_{\alpha_1\alpha_2}={{\partial^2x^\mu}\over{\partial y^{\alpha_1}\partial y^{\alpha_2}}
}\,,\label{3}
\end{equation}

\noindent with obvious extensions to higher order derivatives.

The number of independent components of the metric $g_{\mu\nu}$ is $n(n+1)/2$ and its transformation rule is

\begin{equation}
g_{\alpha\beta}({\bf y})={X^\mu}_\alpha\,{X^\nu}_\beta\,g_{\mu\nu}({\bf x})\,.\label{4}
\end{equation}

\noindent The metric $g_{\mu\nu}$ is a tensor because it transforms like a tensor, that is, the transformation rule involves only $(\partial x/\partial y)$.

For the first derivative of the metric we obtain

\begin{equation}
\partial_\gamma g_{\alpha\beta}({\bf y})=[{X^\mu}_{\gamma\alpha}\,{X^\nu}_\beta+{X^\mu}_\alpha\,
{X^\nu}_{\gamma\beta}]\,g_{\mu\nu}({\bf x})+{X^\mu}_\alpha\,{X^\nu}_\beta\,{X^\lambda}_\gamma\,
\partial_\lambda g_{\mu\nu}({\bf x})\,.\label{5}
\end{equation}

\noindent This is not a tensor since the transformation rule (\ref{5}) involves $(\partial^2x/\partial y^2)$. The number of equations is $n\cdot n(n+1)/2$ while the illegal terms $(\partial^2x/\partial y^2)$ are $n\cdot n(n+1)/2$. Since they are equal we can solve for $(\partial^2x/\partial y^2)$ in terms of $\partial g$. 

For the second derivatives of the metric we obtain

\begin{eqnarray}
\partial_{\delta\gamma}g_{\alpha\beta}({\bf y})&=&[{X^\mu}_{\delta\gamma\alpha}\,{X^
\nu}_\beta+{X^\mu}_\alpha\,{X^\nu}_{\delta\gamma\beta}+{X^\mu}_{\gamma\alpha}\,{X^\nu}_{\delta
\beta}+{X^\mu}_{\delta\alpha}\,{X^\nu}_{\gamma\beta}]\,g_{\mu\nu}({\bf x})\nonumber\\
&&+[({X^\mu}_{\gamma\alpha}\,{X^\nu}_\beta+{X^\mu}_\alpha\,{X^\nu}_{\gamma\beta})\,{X^
\lambda}_\delta+({X^\mu}_{\delta\alpha}\,{X^\nu}_\beta+{X^\mu}_\alpha\,{X^\nu}_{\delta\beta})\,
{X^\lambda}_\gamma\nonumber\\
&&+{X^\mu}_\alpha\,{X^\nu}_\beta\,{X^\lambda}_{\delta\gamma}]\,\partial_\lambda g_{\mu\nu}({
\bf x})\nonumber\\
&&+{X^\mu}_\alpha\,{X^\nu}_\beta\,{X^\lambda}_\gamma\,{X^\rho}_\delta\,\partial_{\rho\lambda}g_{\mu\nu}({\bf x})\,.\label{6}
\end{eqnarray}

\noindent We have $[n(n+1)/2]^2$ equations, $n\cdot n(n+1)(n+2)/6$ illegal terms $(\partial^3x/\partial y^3)$ and $n\cdot n(n+1)(n+2)/6$ illegal terms $(\partial^2x/\partial y^2)$. Second derivatives can be solved from (\ref{5}). For the terms $(\partial^3x/\partial y^3)$ we obtain that the number of relations not involving them is

\begin{equation}
Rie(n)=\left({{n(n+1)}\over2}\right)^2-n\,{{n(n+1)(n+2)}\over6}={{n^2(n^2-1)}\over{12}}\,,
\label{7}
\end{equation}

\noindent which is the number of independent components of the Riemann--Christoffel tensor. In fact, the resulting tensor invariant is the Riemann--Christoffel tensor.

We now generalize the construction above to higher--order derivatives. When considering $d$th--order derivatives the number of equations is

\begin{equation}
E(n,\,d)={{(n+d-1)!}\over{(n-1)!d!}}\,{{n(n+1)}\over2}\,.\label{8}
\end{equation}

\noindent For higher--order derivatives we must take care only of the highest derivatives, since lower ones has already been solved for in the inmediately previous level. The number of illegal terms is

\begin{equation}
U(n,\,d)={{(n+d)!}\over{(n-1)!(d+1)!}}\,n\,.\label{9}
\end{equation}

\noindent Then, the number of relations not involving the derivatives appearing at that order is given by

\begin{equation}
R(n,\,d)=E(n,\,d)-U(n,\,d)={1\over2}\,n\,(d-1)\,{{(n+d-1)!}\over{(n-2)!(d+1)!}}\geq0\,.
\label{10}
\end{equation}

\noindent There are two cases in which the above relation is zero. Firstly, $n=1$, which is trivial since this corresponds to a 1--dimensional space. Secondly, $d=1$, which is the case dealt with in (\ref{5}) above and in which it was shown that we can always solve for $\partial^2x/\partial y^2$. For $d>1$ we must have always more equations than unknowns. In this case we must take care only of the highest derivatives appearing at that level because lower order derivatives have already been solved for in the previous level. 

For $d=2$ eq. (\ref{10}) reduces to

\begin{equation}
Rie(n)=R(n,\,2)={{(n+1)!n}\over{(n-2)!3!2}}={{n^2(n^2-1)}\over{12}}\,,\label{11}
\end{equation}

\noindent which is the number of independent components of the Riemann--Christoffel tensor. 

Scalar invariants are obtained from (\ref{10}) just by substracting the $n(n-1)/2$ conditions we can fix by a local Lorentz transformation. We obtain 

\begin{equation}
S(n,\,d)=R(n,\,d)-{1\over2}\,n\,(n-1)={1\over2}\,n\,(d-1)\,{{(n+d-1)!}\over{(n-2)!(d+1)!}}-{1
\over2}\,n\,(n-1)\,.\label{12}
\end{equation}

\noindent This result coincides with that in \cite{5} for scalar invariants.

Our formulae (\ref{10}) and (\ref{12}) have been obtained by means of a simple counting of equations and illegal terms. In this sense we believe our procedure is clearer than the one used in \cite{5}. Furthermore, our formula (\ref{12}), even when numerically equivalent, is simpler than that in \cite{5}.

In previous works \cite{7}--\cite{9} we have considered a gravitational alternative theory based on a fourth--rank tensor $G_{\mu\nu\lambda\rho}$. The construction of invariants for higher--rank tensors will we reported somewhere else \cite{10}.

\bigskip\bigskip

\centerline{\bf Acknowledgements}

\bigskip

To J. A. Nieto for encouraging me to further pursue the construction of differential invariants. This work was partially supported by Direcci\'on de Investigaci\'on, Sede Bogot\'a, Universidad Nacional de Colombia.

%%%%%%%%%%%%%%%%%%%%%%%%%%%%%%%%%%%%%%%%%%%%%%%%%%%%%%%%%%%%%%%%%%%%%%%%%%%%%%

\end{document}